# Compensation of aberration and speckle noise in quantitative phase imaging using lateral shifting and spiral phase integration

**INHYEOK CHOI,**[1,2] **KYEOREH LEE,**[1,2] **AND YONGKEUN PARK**[1,2,3,*]

[1]*Department of Physics, Korea Advanced Institute of Science and Technology (KAIST), Daejeon 34141, Republic of Korea*
[2]*KAIST Institute for Health Science and Technology, Daejeon 34141, Republic of Korea*
[3]*Tomocube Inc., Daejeon 34051, Republic of Korea*
[*]*yk.park@kaist.ac.kr*

**Abstract:** We present a simple and effective method to eliminate system aberrations and speckle noise in quantitative phase imaging. Using spiral integration, complete information about system aberration is calculated from three laterally shifted phase images. The present method is especially useful when measuring confluent samples in which acquisition of background area is challenging. To demonstrate validity and applicability, we present measurements of various types of samples including microspheres, HeLa cells, and mouse brain tissue. Working conditions and limitations are systematically analyzed and discussed.

## 1. Introduction

Quantitative phase imaging (QPI) is a rapidly growing technique for imaging biological samples [1, 2]. Employing the principle of interferometric imaging, QPI measures the phase retardation induced by a sample, from which 2D phase delay maps or 3D refractive index (RI) tomograms of the sample can be retrieved. Due to its non-destructive, label-free, and quantitative imaging capability [3], QPI has been widely used for study in various biological and medical fields, including hematology [4, 5], immunology [6], neuroscience [7-9], cell biology [10, 11], microbiology [12], histopathology [13, 14], and nanotechnology [15-17].

The imaging quality in QPI is determined by speckle noise and aberration. Speckle noise, unwanted diffraction patterns due to the use of coherent illumination and multiple reflection or scattering from the surface of optical elements or dust particles, can be alleviated by various approaches. For example, the use of temporally or spatially incoherent illumination decreases the effects of speckle noise in QPI [18-21] or common-path interferometric geometry [22-25].

Compensation for system aberration has frequently been performed by the background subtraction method [26]; the aberration in a measured sample field is numerically subtracted using a field image obtained from a background or no-sample region under the same optical setup. Although the background subtraction method has been widely used, measurements of clear background regions are often not available in practical situations, especially for confluent biological cells and tissue slices. This constraint is unfortunate because QPI, with its unique label-free contrast modalities, its quantitative imaging capability, and its avoidance of damage to samples, has much to offer the fields of cell biology and histopathology. Alternatively, parametric approaches have been introduced in which system aberrations are simply modeled with a few parameters [27-30]. However, these approaches have limitations, especially when high orders of aberration are involved.

Here, we propose a simple but general method for the compensation of aberration and speckle noise. By exploiting the linear shift invariant property, speckle noise and aberration terms are decoupled from captured holograms with lateral shifts. Three holograms are captured while laterally shifting the samples in orthogonal directions, and by the static parts – the speckle noise and aberration – are cancelled each other out. The phase integration method is utilized to reconstructed introduced the speckle- and aberration-free phase image from the rest differential information [31]. The applicability of the present method is demonstrated using various samples including polystyrene microspheres, eukaryotic cells, and a mouse brain tissue slice. Systematic analyses of working conditions and limitations are also discussed.

## 2. Principles

Figure 1 depicts the working principles of the present method. The quantitative phase image $\phi(x, y)$ retrieved from a hologram can be decomposed into an ideal aberration-free phase image $\phi_0(x, y)$ and an aberration term $\varepsilon(x, y)$ as $\phi(x, y) = \phi_0(x, y) + \varepsilon(x, y)$, as shown in Figs. 1(a)−(b). In the background subtraction method [26], aberration term $\varepsilon(x, y)$ is obtained by measuring the background hologram of an area without a sample [Fig. 1(c)]. Then, $\varepsilon(x, y)$ is subtracted from the measured $\phi(x, y)$ in order to retrieve $\phi_0(x, y)$, [Fig. 1(d)]. However, the background subtraction method has difficulty in finding clean background regions, especially for confluent cells or tissue slices.

The present method does not require finding a background region to measure $\varepsilon(x, y)$. Instead, the method precisely retrieves $\varepsilon(x, y)$ from three holograms with lateral shifts even in the presence of confluent samples. One hologram of a sample is recorded, and two other holograms are recorded after shifting the sample in orthogonal directions. From three measured holograms, original and shifted phase images are obtained. Subtracting the original phase image from shifted phase images, two differential phase images $\Delta_x\phi(x, y)$ and $\Delta_y\phi(x, y)$ are calculated as $\Delta_x\phi(x, y) = \phi_0(x + \Delta x, y) - \phi_0(x, y)$ and $\Delta_y\phi(x, y) = \phi_0(x, y + \Delta y) - \phi_0(x, y)$, respectively, [Figs. 1(e)−(f)].

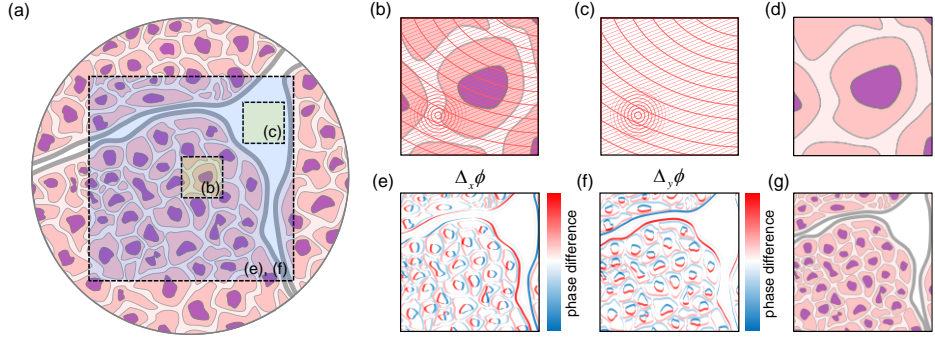

**Fig. 1 Principles of shift differential method.** (**a**) In conventional background subtraction method, a sample region (**b**) and a background region (**c**) are recorded. In shift differential method, sample region and its vertical and horizontal shifts are recorded. (**b**) Phase image of a sample region. (**c**) Aberration image obtained from background phase image. (**d**) Improved phase image, which is the subtraction of the aberration image from the phase image. (**e**) Horizontal shift differential phase image $\Delta_x\phi(x, y)$. (**f**) Vertical shift differential phase image $\Delta_y\phi(x, y)$. (**g**) Phase image retrieved from spiral phase integration of differential phase images.

In order to retrieve the ideal phase image from two differential phase images, we utilized spiral phase integration [31]. According to the Fourier shift theorem, the Fourier transforms of $\Delta_x\phi(x, y)$ and $\Delta_y\phi(x, y)$ are $(e^{2\pi i \Delta x u} - 1)\Phi_0(u,v)$ and $(e^{2\pi i \Delta y v} - 1)\Phi_0(u,v)$, respectively, where $\Phi_0(u,v) = FT[\phi_0(x, y)]$. Next, we define $G(u,v)$ and $H(u,v)$ as follows,

$$G(u,v) = FT\left[\Delta_x\phi(x,y) + i\Delta_y\phi(x,y)\right] = H(u,v)\Phi_0(u,v), \qquad (1)$$

where

$$H(u,v) = e^{2\pi i \Delta x u} + i e^{2\pi i \Delta y v} - 1 - i. \qquad (2)$$

Dividing $G(u,v)$ by $H(u,v)$, we obtain $\Phi_0(u,v)$ and subsequently the ideal phase image $\phi_0(x, y)$. See Appendix A for detailed derivations.

## 3. Method

### 3.1. Optical setup

The experimental setup is presented in Fig. 2. A coherent plane-wave beam from an He-Ne laser ($\lambda$ = 633 nm, HNL050R, Thorlabs Inc.) is split by a beam splitter into a reference and a sample beam. Samples were mounted on a motorized scanning stage (MLS203-1, Thorlabs Inc.) for the automatic shift. The light diffracted from the samples was collected by a high numerical aperture (NA) objective lens (NA=1.2, water immersion, UPLSAPO 60XW, 60×, Olympus, Inc., Japan). The sample beam was further magnified by a factor of two and interfered with a reference beam at the image plane. Interference pattern was recorded using a complementary metal–oxide semiconductor camera (DCC3240M, Thorlabs Inc.).

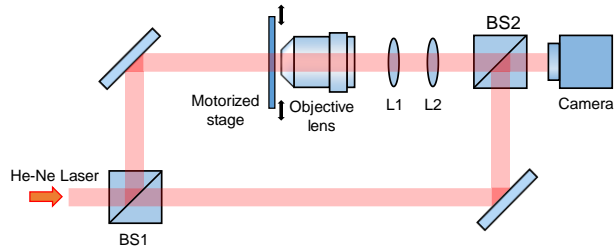

**Fig. 2 Optical Setup.** Sample is mounted on a motorized stage. Sample and reference beam interfere and generate a hologram, which is recorded by the camera.

### 3.2. Sample preparation

Silica (SiO$_2$) beads ($n$ = 1.4570 at $\lambda$ = 633 nm) with diameter of 3 μm were immersed in water and sandwiched by two cover slips. Polystyrene beads ($n$ = 1.5875 at $\lambda$ = 633 nm) with a diameter of 10 μm were immersed in index matching oil ($n$ = 1.5279 at $\lambda$ = 633 nm) and sandwiched by two cover slips. HeLa cells (human cervical cancer cell line) were cultured in DMEM (Dulbecco's modified Eagle's media) with 10% of FBS (fetal bovine serum) and 1% of penicillin-streptomycin, and fixed with

4% formaldehyde. Fixed cells were prepared in a Petri dish and immersed in PBS (Phosphate-buffered Saline). Mouse brain tissue was obtained from a 22-year-old male mouse. After fixation and dehydration, brain tissue was sliced and sandwiched between two cover slips with a mounting medium ($n$ = 1.355).

## 4. Results

To demonstrate the validity of the present method, we captured phase images of the 3-μm-diameter silica beads. Phase images obtained by background subtraction method and by the present method are presented in Fig. 3. Figure 3(a) displays a raw phase image $\phi(x, y)$ retrieved from a single hologram. In the background subtraction method, an aberration phase image $\varepsilon(x, y)$ is additionally measured. Then, $\varepsilon(x, y)$ was subtracted from $\phi(x, y)$ to obtain an improved phase image $\phi_0(x, y)$, as presented in Fig. 3(c).

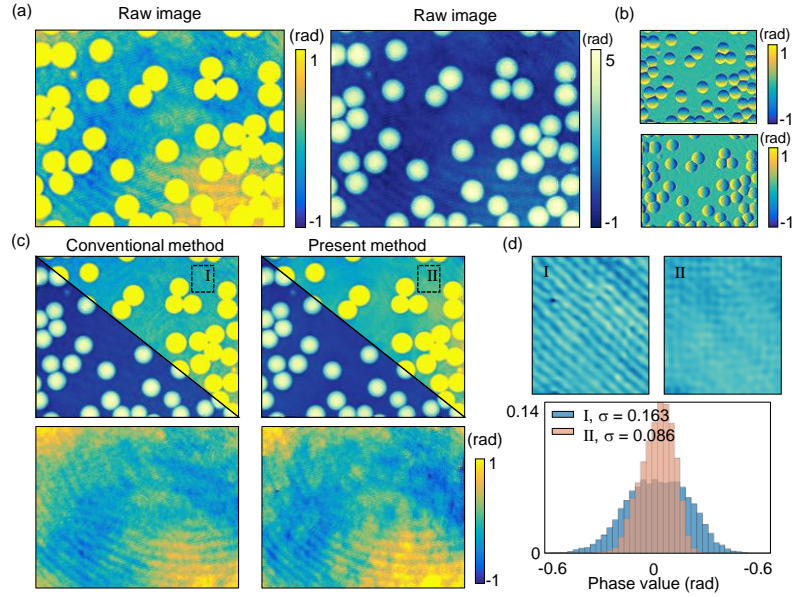

**Fig. 3 Proof-of-principle demonstration with silica beads.** (**a**) Raw phase images retrieved from a measured hologram (displayed in two color scales). (**b**) Differential phase images with vertical and horizontal shifts. (**c**) First row: Phase images improved by conventional background subtraction method and by the present method. Second row: aberration images retrieved from each method. (**d**) Phase map of regions I and II in the reconstructed images, and their phase histograms.

In contrast, the present method utilizes two shift differential phase images $\Delta_x\phi(x, y)$ and $\Delta_y\phi(x, y)$ [Fig. 3(b)]. Spiral phase integration of the differential phase images yields the reconstructed phase image $\phi_0(x, y)$, in which the aberration phase term $\varepsilon(x, y)$ is automatically canceled out. Based on the reconstructed phase image, we can also retrieve $\varepsilon(x, y)$ by subtracting the reconstructed phase image from the raw phase image [Fig. 3(c)].

The phase image improved via the conventional method involves high-frequency aberration that originates from a slight variation of the angle between the sample and the reference beam, as shown in Fig. 3(d). On the other hand, the phase image improved by the present method showed relatively cleaner phase images. The standard deviation of the phase values in the flat no-sample area are 0.163 and 0.086 rad for the conventional method and the present method, respectively.

To demonstrate the applicability of the present method to biological samples, phase images of HeLa cells are measured [Fig. 4]. In the raw phase images, due to the system aberration, the quality of phase images is severely deteriorated [Fig. 4(a)]. In contrast, when the present method is applied, the background phase is removed and, then, the details of the sample can be clearly seen in the phase images [Fig. 4(b)]. The overall quality of the aberration-corrected images when using the present method is comparable to that of images measured with white-light QPI techniques [32-34].

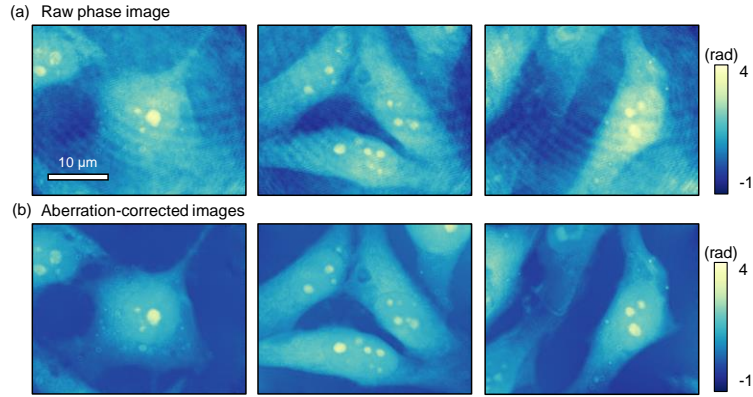

**Fig. 4 Phase images of HeLa cells.** (**a**) Raw phase images of HeLa cells. Aberrations are present. (**b**) Phase images of HeLa cells improved by present method. Aberrations are clearly removed, and details of the sample can be seen in the phase images.

To further demonstrate the potential of the present method, a phase image of mouse brain tissue slice is presented [Fig. 5]. 25 phase images were measured and aberration was corrected with the present method. Then, the 25 phase images were manually stitched together to construct a large field-of-view phase image [Fig. 5(a)]. Representative raw and improved phase images are presented in Figs. 5(b)–(c). Raw phase images contain significant aberration artifacts, which can clearly be removed using the present method.

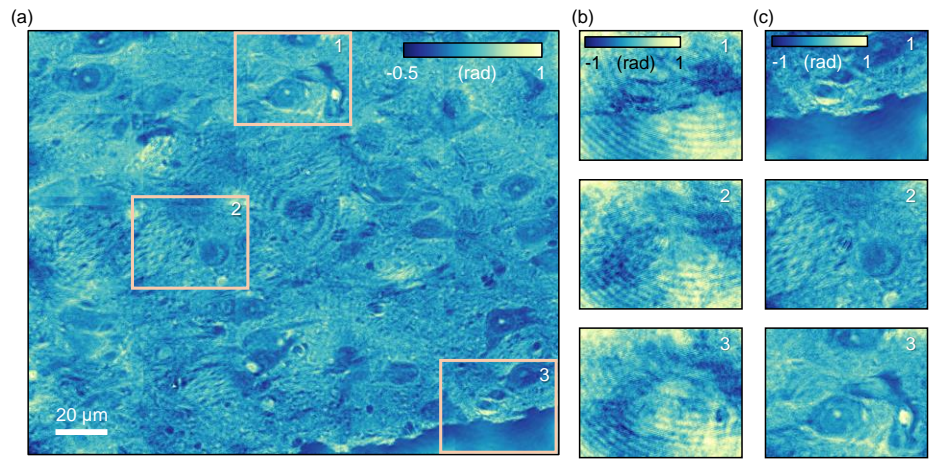

**Fig. 5 Phase images of large tissue slide.** (**a**) Mouse brain tissue phase image; using the present method, this image was numerically combined from 25 phase images. (**b**) Raw phase images of the mouse brain tissue, in subareas denoted 1, 2, and 3. (**c**) Corresponding aberration-corrected phase images obtained using the present method.

## 5. Discussion and summary

In this Letter, exploiting the spiral phase integration of differential phase images, we propose and experimentally demonstrate a method of compensation for aberration and speckle noise in QPI. This method does not require the measurement of any sample region, which limits the applicability of QPI for confluent samples such as high-density cell cultures and tissue slices. Instead, the present method measures three laterally translated phase images in the presence of samples, from which only set of sample phase information can be reconstructed; other static noises such as system aberration and speckle patterns are systematically canceled out.

We demonstrate that the present method effectively removes aberrations from various types of samples including silica beads, eukaryotic cells, and tissue slices. We also compare the performance of the present method to that of the conventional background subtraction method. Although both methods successfully remove aberration, remaining error terms differ. The background subtraction method is susceptible to physical vibration or the alteration of the path geometry during the measurement of the no-sample region, leaving aberrations of high spatial frequency from multiple reflections. However, the present method removes most of the aberration, including multiple light scattering, providing two-fold enhancement in phase sensitivity. Nonetheless, the present method produces artifacts of low spatial frequency at the point of abrupt phase change. However, this issue can be alleviated by using a high NA objective lens or by reducing abrupt phase changes by matching phase contrast between a sample and a medium.

As shown in Figs. 4 and 5, the present method is also applicable to samples with complex contour, including cells and biological tissues. Moreover, phase images corrected with the present method can be successfully stitched together to reconstruct large field-of-view phase images. We also verify that our method is applicable to phase images that occupy the boundary of the field-of-view, including tissue samples.

One of the limitations of the present method is that it requires the measurement of two additional phase images with lateral shifts in orthogonal directions. However, using a motorized stage, these additional measurements can be performed in a short period of time. Because the present method is not limited by the type of instrumentation but is generally applicable to optical field measurement techniques, it can also be readily applied to various types of setup, ranging from digital holographic microscopy [35, 36] to optical diffraction tomography [37, 38].

In summary, we present a simple but powerful method of compensation for aberration and speckle noise in QPI. The present method will be applied to various fields, in particular the study of as confluent samples such as adherent biological cells and tissue slices.

## Funding

This work was supported by KAIST, BK21+ program, Tomocube, and the National Research Foundation of Korea (2015R1A3A2066550, 2014M3C1A3052567, 2014K1A3A1A09-063027).

## Acknowledgement

We thank Dr. Kyoohyun Kim and Mr. Jaehwang Jung for helpful discussion regarding the experiment and simulation analyses, and Mr. Sangyun Lee for the image visualization. Mr. Shin, Mr. Lee, and Prof. Park have financial interests in Tomocube Inc., a company that commercializes optical diffraction tomography and quantitative phase imaging instruments and is one of the sponsors of the work.

## References and links

## Appendix

*1. The formulation of the algorithm*

Spiral phase integration was first introduced in Ref. [31] for differential interference contrast (DIC) microscopy. It has been widely applied for DIC microscopy [39] and quadriwave lateral shearing interferometry (QWLSI) [40, 41]. The spatial phase integration assumes spatial periodicity in an image, because the shift theorem for finite Fourier transform is applied to circular shift to an image. When a differential phase image does not satisfy spatial periodicity, artifacts occur at the boundary of the image, as described in [31, 41]. In this Letter, boundary parts of shift differential phase images $\Delta_x \phi(x, y)$ and $\Delta_y \phi(x, y)$ were modified to match spatial periodicity and prevent boundary artifacts.

During spiral phase integration, the division of $G(u,v)$ by $H(u,v)$ generates singularities at $H(u_c, v_c) = 0$, which occurs when

$$(u_c, v_c) = \left( \frac{n}{\Delta x}, \frac{m}{\Delta y} \right) \text{ or } \left( \frac{n+\frac{1}{4}}{\Delta x}, \frac{m+\frac{1}{4}}{\Delta y} \right) \tag{3}$$

for integers $n$ and $m$. To remove singularities of the latter case, we defined $G'(u,v)$ and $H'(u,v)$ by the phase shift of $\pi$, as follows,

$$G'(u,v) = \left( e^{2\pi i \Delta x u} + i e^{2\pi i \Delta y v} - 1 + i \right) \Phi_0(u,v), \tag{4}$$

where

$$H'(u,v) = e^{2\pi i \Delta x u} + i e^{2\pi i \Delta y v} - 1 + i, \tag{5}$$

and calculated the division of $G'(u,v)$ by $H'(u,v)$ and added two results with the weight of $|H(u,v)|$ and $|H'(u,v)|$, respectively. In order words, we calculated

$$\Phi_0(u,v) = \frac{1}{|H(u,v)| + |H'(u,v)|} \left( \frac{|H(u,v)|}{H(u,v)} G(u,v) + \frac{|H'(u,v)|}{H'(u,v)} G'(u,v) \right), \tag{6}$$

*2. The effects of the shift size*

The effects of the shift size on the quality of image improvement are investigated. Figure 6(a) presents a raw phase image and a phase image improved by the background subtraction. To test the effects of the shift size, various shift sizes were applied to the present method. As displayed in Fig. 6(b), the best improvement is achieved when the shift is comparable to the diffraction limit size of the system (527.5 nm). When the shift was smaller than the diffraction-limited size, low-frequency aberration and artifacts are present whereas the shift is greater than the diffraction limited size, high spatial frequency artifacts arise near samples.

To investigate the effects of the shift to the image quality, Fourier spectra were shown. The second row of Fig. 6(c) presents the intensity map of $|H(u,v)|+|H'(u,v)|$, and the third row presents the retrieved phase spectra in Fourier domain $\Phi_0(u,v)$. When the shift size is smaller than the diffraction limit, $|H(u,v)|+|H'(u,v)|$ attains small values near the center of the Fourier plane. In this case, division by $H(u,v)$ and $H'(u,v)$ leads to the exaggeration of the low-spatial-frequency errors, as shown in the reconstruction image. In contrast, when the shift size is larger than the diffraction limit, zeroes of $|H(u,v)|+|H'(u,v)|$ other than $(u,v) = (0,0)$ located inside the NA circle (the dotted circle in Fig. 6). Thus, some of the high-spatial-frequency components of the image are lost, as shown in the reconstruction image.

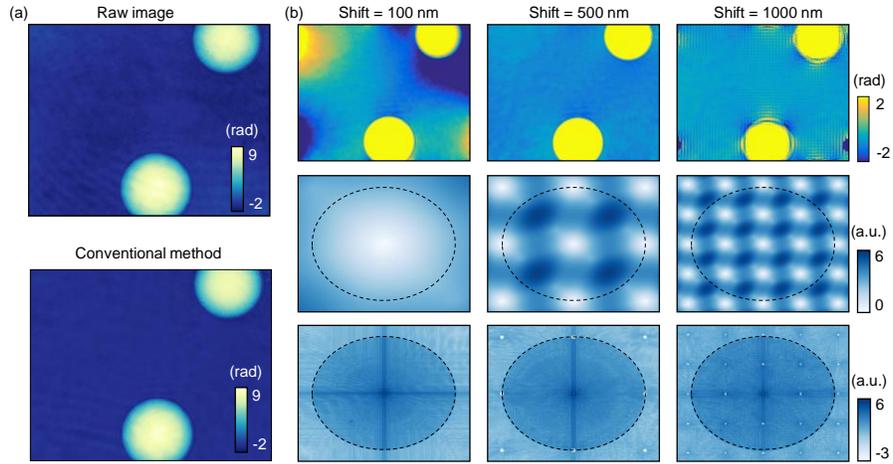

Fig. 6. Effects of shift size on the image reconstruction. Polystyrene beads with the diameter of 10 μm are imaged. (a) Raw phase image and its improvement using the background subtraction method. (b) Phase images are improved using the present method with shift sizes of 100, 500, 1000 nm. $|H(u,v)|+|H'(u,v)|$ and $\Phi_0(u,v)$ are presented in the second and the third rows, respectively. In the third rows, the area inside the NA circle is emphasized by addition of constant values.